\documentclass[conference]{IEEEtran}
\IEEEoverridecommandlockouts
\usepackage{cite}
\usepackage{amsmath,amssymb,amsfonts}
\usepackage{algorithmic}
\usepackage{graphicx}
\usepackage{textcomp}
\usepackage{xcolor}

\def\BibTeX{{\rm B\kern-.05em{\sc i\kern-.025em b}\kern-.08em
    T\kern-.1667em\lower.7ex\hbox{E}\kern-.125emX}}
\begin{document}

\title{Attention-based Region of Interest (ROI) Detection for Speech Emotion Recognition}

\author{\IEEEauthorblockN{Jay Desai}
\IEEEauthorblockA{\textit{Department of Computer Science} \\
\textit{New York Institute of Technology}\\
New York, USA \\
jdesai09@nyit.edu}
\and
\IEEEauthorblockN{Houwei Cao}
\IEEEauthorblockA{\textit{Department of Computer Science} \\
\textit{New York Institute of Technology}\\
New York, USA \\
hcao02@nyit.edu}
\and
\IEEEauthorblockN{Ravi Shah}
\IEEEauthorblockA{\textit{Department of Computer Science} \\
\textit{New York Institute of Technology}\\
New York, USA \\
rshah79@nyit.edu}
}

\maketitle

\begin{abstract}
Automatic emotion recognition for real-life applications is a challenging task. Human emotion expressions are subtle, and can be conveyed by a combination of several emotions. In most existing emotion recognition studies, each audio utterance/video clip is labelled/classified in its entirety. However, utterance/clip-level labelling and classification can be too coarse to capture the subtle intra-utterance/clip temporal dynamics. For example, an utterance/video clip usually contains only a few emotion-salient regions and many emotionless regions. In this study, we propose to use attention mechanism in deep recurrent neural networks to detection the Regions-of-Interest (ROI) that are more emotionally salient in human emotional speech/video, and further estimate the temporal emotion dynamics by aggregating those emotionally salient regions-of-interest. We compare the ROI from audio and video and analyse them. We compare the performance of the proposed attention networks with the state-of-the-art LSTM models on multi-class classification task of recognizing six basic human emotions, and the proposed attention models exhibit significantly better performance. Furthermore, the attention weight distribution can be used to interpret how an utterance can be expressed as a mixture of possible emotions.
\end{abstract}


\section{Introduction}
Human emotion plays an important role in everyday decision making. It impacts the way they communicate i.e. via body language, facial expressions, verbal communication, personality, etc\cite{b1,b2} and also shows the characteristics of the person like\cite{b3} leadership, honesty, teamwork. In fact, in a book Emotional Intelligence 2.0 by Travis Bradberry, Ph.D. it was found that 90\% of all the people they studied at work had higher EQs. Speech is the key aspect of expressing emotion and most basic way of human-human and human-machine interactions. In this study, we use the acoustic features of speech to detect the part/region of speech which impacts the most in detecting the emotional state of the subject.\par
Speech emotion recognition (SER) is a trending topic in the field of research with the rapid use of artificial intelligence and machine learning in our lives. While there are many techniques and features that has been proposed\cite{b4,b5} but it is still not clear on which gives most information about emotions. There are two common types of feature extraction methods, Global features and Local features. Features extracted for whole audio file using full audio clip are usually known as global features while features extracted using multiple time frames at regular intervals usually 20-30ms with or without overlap are known as local features. Global features usually work well for general machine learning models whereas local features are better suited for deep neural networks. Features are the Low Level Descriptors (LLDs) which are believed to affect the most to emotions and High Level Statistical Functions (HSFs) which are applied to LLDs to extract variations and contours for temporal description of the data. While the majority of researchers have agreed that global features perform better than local features\cite{b6,b7,b8,b9} but other researchers also believe that global features only show better performance for high arousal emotions like anger, fear, joy\cite{b10}. Here are some common LLDs and HSFs:
\begin{table}[h]
\caption{Common LLDs and HSFs\cite{b11,b12,b13}}
\begin{center}
\begin{tabular}{p{4cm} p{4cm}}
\hline
\textbf{LLDs} & \textbf{HSFs} \\
\hline
pitch, MFCCS, voicing probability, energy, Zero-Crossing rate, formant locations/ bandwidths, harmonics-to-noise ratio, jitter, loudness, etc. &
mean, variance, min, max, range, median, quartiles, skewness, kurtosis, linear regression coefficients, RMSenergy, SMA, PCM, Rfilters, etc.\\
\hline
\end{tabular}
\label{tab1}
\end{center}
\end{table}

\begin{figure*}[!t]
\includegraphics[width=\textwidth]{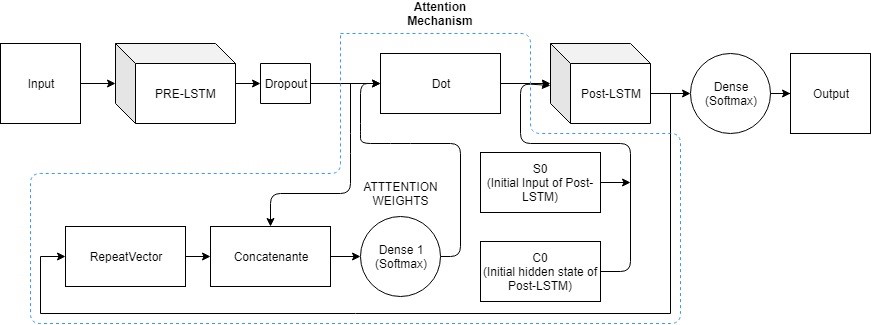}
\caption{Architecture for Model 1 LSTM with Attention (the dotted line shows the attention mechanism block).}
\label{arch}
\end{figure*}
\par
In recent years, there has been advancements in the field of machine learning and artificial intelligence techniques and their applications. The authors in \cite{b14,b15,b16,b17,b18} used ensemble of SVM instead of just one to better classify emotions using MFCC, total energy and f0 features from audio signal. The work done in \cite{b19,b20} focused on using Deep Neural Networks and Recurrent Neural Network (RNNs) to learn short term acoustic features at frame level and then mapping them to sentence-level representation using extreme learning machines (ELM). The authors in \cite{b21} used various LSTM techniques and used logistic regression based general attention mechanism with weighted pooling for classification.\par
The authors in \cite{b22} used deep convolutional recurrent neural networks where they used a combination of convolutional neural network (CNN) layers and bi-directional LSTM layers. They derived a generalized attention mechanism method from the methods proposed in \cite{b23,b24} and compare CNNs task-specific spectral decorrelation with that of the discrete cosine transformation (DCT) in clean and noisy conditions.
One issue that appears in most SER datasets is their labels are given at utterance level and in any speech, there are many silence periods and only a short time of utterance of few words that impact the overall emotion. The silent frames can be handled by labelling them as null, our approach handles it implicitly without any separate or explicit mechanisms to handle it.\par
In this paper, we used an LSTMs in an encoder - decoder based fashion proposed in \cite{b25,b26} using LSTM. Different from the attention technique used in \cite{b21} where they used logistic regression and mean pooling, in our approach for attention, we train a separate small Neural Network to learn how much attention to pay on each frame. Our method works similar to human mechanism for translating the input signals in brain and using only contextually important parts to decode the input and return the results as emotions. This has been proven better in many language translation tasks where to translate one word, only a part of the sentence is important rather than whole sentence. In the following we discuss various approaches we used for classification.

\section{Neural Network Model}
In the below subsections data description describing which dataset is used and details about it, features extraction techniques describing all the features extracted tools or libraries used, testing strategy, various models have been described.

\subsection{Data Description}
The dataset used here is Crowd-sourced Emotional Multimodal Actors Dataset (CREMA-D) \cite{b27}. It is a dataset of 7442 audio files from 91 different subjects - 48 males and 43 females of various races such as African, Asian, American, Hispanic, Caucasian between the ages of 20 and 74. Subjects spoke 12 sentences in 6 different emotions i.e. Anger, Disgust, Fear, Happy, Neutral and Sad in four emotion levels - low, medium, high and unspecified. Categorical emotion labels and real-value intensity values for the perceived emotion were collected using crowd-sourcing from 2,443 raters. All 91 actors read 12 sentences in three to four emotion levels making total number of files to be 7442 clips. Each actor has an average of 82 audio clips. The problems with other datasets such as low recognition rates of human subjects\cite{b10}, bad quality of recorded utterances is not found in this dataset. The human recognition rates for intended emotion for audio only data was 40\%. Recognition rates were highest for neutral followed by happy, anger, disgust, fear and sad.

\subsection{Feature Extraction}
Every audio clip has been padded with zeros till maximum audio clip length to remove the inconsistencies between audio clip length. The first 13 Mel-frequency cepstral coefficients are extracted for every 20 milliseconds (frame length) with 50\% or 10 milliseconds (frame step) of overlap between each frame for the data used in model 1,2,3 and 4.

\subsection{Testing Strategy}
We have used Leave One Subject Out (LOSO) strategy to train-test-split the data. In that for every 91 subjects that a model gets trained on, it gets tested on one subject. The total training files for one model comprised of 7360 training files and testing on one subject having 82 files. In total of 92 models were trained and tested and for the final result, mean from all the results from 91 models was taken and a confusion matrix was created as shown in Fig.~\ref{cm}. It gives a better information on how the model will perform on real world scenario because it has never seen the subject before than other methods such as k-fold cross validation or random train-test split where the model has already seen and trained on the audio clips of subject and can perform better on them in testing phase.

\subsection{Model 1 - LSTM with Attention }
We have used a Sequence to Sequence type model which has following parts:
\begin{enumerate}
  \item Encoder (Pre-LSTM)
  \item Intermediate Attention Layer environment
  \item Decoder (Post-LSTM)
  \item Dense (SoftMax) Layer
\end{enumerate}
\par
The Pre-LSTM and Post-LSTM layers are CuDNNLSTM and are unidirectional. The output of Pre-LSTM is given to the attention mechanism layer explained below which returns attention weights. These attention weights are dotted with the output of Pre-LSTM and fed to Post-LSTM followed by dense or fully connected layer with softmax activation function which calculates the probability for 6 classes. Dropout is used to reduce overfitting for regularization.\par
In general, LSTMs have ability to learn from the sequence of data and also learn the dependencies between sequences and store them in the memory cell for future use. It stores the relevant information and forgets the irrelevant information using the forget gate and passes on this learnt context to next time steps. The context learnt here is the result of backpropagation and loss minimization to optimize the overall accuracy. The attention mechanism used in model 1 and model 2 is described and explained below.\par
Attention Mechanism: The attention block shown in figure ~\ref{arch} shows the architecture of attention mechanism in the dotted line. Models without attention passes the output from dropout layer directly to post-LSTM layer. The similar attention mechanism is used in \cite{a,b} for generating image captions while focusing on only parts of the image.\par
There are two types of LSTM layers used for pre-LSTM, one is uni-directional and other is bi-directional. The post-LSTM passes outputs $o<t>$ and hidden cell state $h<t>$ from one time step to next. The inputs to post-LSTM are $s<t>$ , context and $h<t>$. The outputs of pre-LSTM are represented as $p<t>$ for unidirectional and for bidirectional LSTM, the forward and backward direction, the outputs are concatenated $p<t> = [ p<t>(forward),p<t>(backward) ]$. The repeatVector copies $o<t-1>$ for $x$ times where $x$ is the number of time frames used to extract the data and then Concatenate layer concatenates it with $p<t>$ to compute $e<t,t’>$ which is then passed to dense layer and then softmax layer to output $a<t,t’>$ where the $t$ represents the post-LSTM’s time step and $t’$ represents the pre-LSTM’s time step.\par
At any time step $t$, given the outputs of pre-LSTM $[p<1>,p<2>,…,p<x>]$ and the previous output of post-LSTM $o<t-1>$, the attention mechanism will compute attention vector or attention weights $[a<t,1>,a<t,2>…a<t,x>]$ and output the context vector shown in (1)
\begin{equation}
\label{eq:1}
\begin{split}
context<t> &= \sum_{t'}^{x} a<t,t'>p<t'> 
\end{split}
\end{equation}

\subsection{Model 2 - Bidirectional LSTM with Attention}
A basic LSTM cell has memory cell which preserves information for learning Long Term Dependencies. It learns the dependencies of the inputs that are passed through it and uses that for the next output. A Uni-Directional LSTM only preserves past information where the inputs are passed forward as they come whereas the bidirectional LSTM passes the inputs in both forward and backward directions. For instance, let’s say on a high level the LSTM predicts the emotion for various time steps with inputs given in forward and inputs given in backward direction as shown below,\\
Forward LSTM: Angry, Sad, Angry, Fear, Angry….\\
Backward LSTM: …. Fear, Fear, Fear, Fear, Fear, Fear, Fear.\par
Here one can see that the past and future information both can make it easier for model to predict the emotion for next time step.\par
The hidden state in bi-directional LSTM is the concatenation of hidden states from forward run and backward run of the inputs. Bi-directional LSTM generally outperforms the uni-directional LSTM and it is becoming a new standard where the input sequence doesn’t matter. In this model we have changed the type from uni-directional to bi-directional in Pre-LSTM layer of model 1.

\subsection{Model 3 - LSTM without attention}
The model consists of 4 layers: Pre-LSTM (uni-directional), Dropout, Post-LSTM, Dense. The model is almost like model 1 but without the attention mechanism. We removed the attention mechanism in this model to compare the results with model 1 and check the impact attention makes on the results. The model architecture can be seen in Fig.~\ref{arch} by removing the attention block. The output of Pre-LSTM is fed to the dropout layer followed by Post-LSTM. Then the final layer is dense or fully connected layer with softmax as activation function.

\subsection{Model 4 - Bidirectional LSTM without attention}
This model is same as model 3 except for the Pre-LSTM layer is replaced by Bidirectional Pre-LSTM layer whose outputs are combined and fed to next layer.

\begin{figure}[t]
\centerline{\includegraphics[ scale=.675 ]{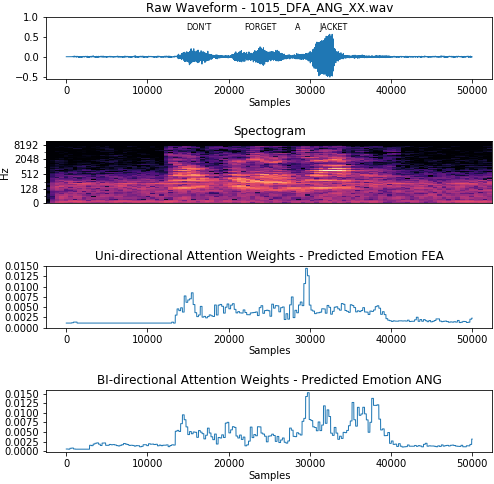}}
\caption{For audio file 1015\_DFA\_ANG\_XX.wav, RAW Waveform, Spectrogram, Uni-Directional Attention Weights (Model 1) and Bi-directional Attention weights in order are shown.}
\label{spec}
\end{figure}

\begin{figure*}[!t]
\includegraphics[width=\textwidth]{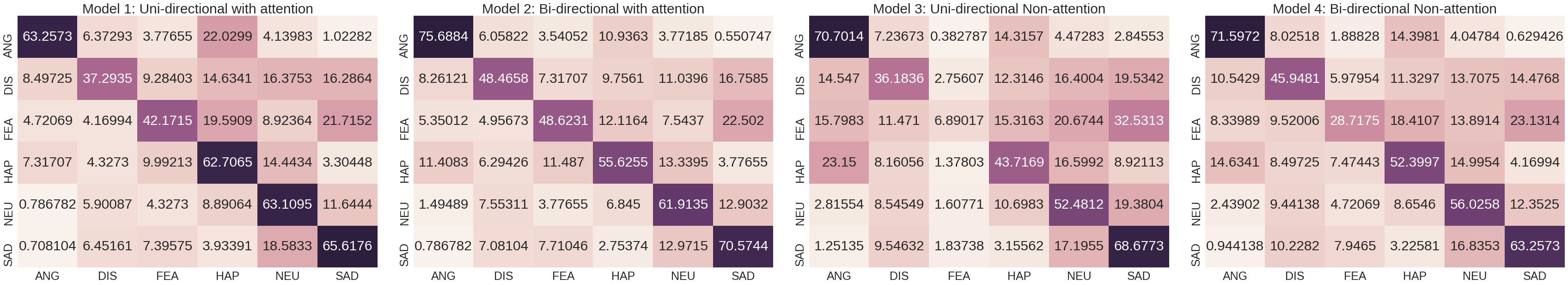}
\caption{Mean normalized confusion matrices of 91 subjects tested on 91 models trained using LOSO strategy. UNI: Uni-directional LSTM. BI: Bi-directional LSTM. NA: Non (without) -Attention model. AT: Attention.}
\label{cm}
\end{figure*}

\section{Evaluation and Result Analysis}
We compare the results of model 1 and model 2 to analyse the attention values for unidirectional LSTM and bidirectional LSTM. For example, in a statement ‘DFA’ recorded by subject 1015 in ‘ANG’ anger emotion with unspecified emotion level, attention weights from model 1 and model 2 are shown with spectrogram and raw waveform above them. It can be observed that both the models can handle the silent areas very well and the values tend to zero. It can also be seen that Bi-Directional LSTM helps the model predict the correct emotion compared to Uni-Directional LSTM at the word ‘JACKET’ which spans from 30000th sample to 40000th sample which can be seen in Fig.~\ref{spec}. Note that the values for attention are the output of a softmax function which gives the probability for every time frame and they have been expanded to fit the samples in the raw waveform and spectrogram shown in figure 3.\par
We used confusion matrix to describe the performance of the classification model on test data for each model. For 91 models, 91 confusion matrices were generated, aggregated and normalized to generate one final confusion matrix which represents the overall model performance.  Each item(I,j) in confusion matrix tells the number of items belonging to emotion I classified as emotion J. 
\begin{table}[h]
\caption{The best and worst performing models on various emotions showing the (\%) of true positives.}
\begin{center}
\begin{tabular}{p{2cm} p{3cm} p{2.6cm}}
\hline 
\textbf{Emotion} & \textbf{Best Classified (\%)} &  \textbf{Worst Classified (\%)} \\
\hline
Anger	& Model 2 – 75.6	& Model 1 – 63.25 \\
Disgust	& Model 2 – 48.46	& Model 3 – 36.18 \\
Fear	& Model 2 – 48.62	& Model 3 – 6.89  \\
Happy	& Model 1 – 62.7	& Model 3 – 43.71 \\
Neutral	& Model 1 – 63.10	& Model 3 – 52.48 \\
Sad		& Model 2 – 70.57	& Model 4 -- 63.25\\
\hline
\end{tabular}
\label{tab2}
\end{center}
\end{table}

\par
As seen in table~\ref{tab2}, model 2 best classifies the anger, disgust, fear and sad emotions of all the models. Model 1 best performs on happy and neutral emotions and worst performs on anger emotion of all the models. Model 3 worst performs on disgust,
 fear, happy and neutral emotions of all the model. Model 4 worst performs on sad emotion of all the models. From this it can be deduced that model 3 is the worst performer and model 4 is the best performer.

\subsection{Uni-directional Vs Bi-directional Results}
Comparing the results of model 1 and model 2, we can see that model 2 classifies every emotion better than model 1 except for “happy” emotion where model 1 has 62.6\% true positives vs 55.62\% of model 2. Comparing the results of model 3 and model 4, we can see that model 4 outperforms model 3 for all emotions except for emotion “sad” having 68.67\% true positives vs 63.25\% true positives. This shows that the inputs passed to LSTM in both the directions help the model to perform better at classifying compared to the inputs passed only in one direction.

\subsection{Attention vs Non-Attention}
Comparing the unidirectional results - model 1 and model 3, model 1 with attention performs better than model 3 for emotions: disgust, neutral, happy and significantly better in classifying fear.  Comparing the bidirectional results – model 2 and model 4, model 2 with attention outperforms model 4 in every emotion. Although model 4 was able to classify the fear emotion better due to bidirectional LSTM than model 3, adding attention to it model 2 performs better than all of the models in classifying fear emotion.\par
Finally, out of all the four models, model 2 with Bidirectional LSTM and Attention mechanism outperforms every other model.

\section{Discussion}
From the results of all the experiments on the Crema-D dataset and using the LOSO strategy and extracting 13 mfcc features for 20ms time-frame and 10ms time-step, we conclude that using a bi-directional LSTM instead of uni-directional LSTM helps the model classify the emotions better and using attention mechanism improves the performance and helps handle the noisy, silent and other non-useful parts of speech. However, increasing the number of input features along with 13 mfccs such as pitch, total energy, mean, variance, etc. may contribute to better performance but may significantly increase the training time and use far more resources in terms of gpu, memory bandwidth, CPU, storage, etc. Furthermore, multiple classifiers as seen in \cite{b28} can be combined as shown in \cite{b29,30} in hierarchical, serial and parallel fashion.

\section{Conclusion}
In future, the more complex feature sets can be used such as LLDs and HSFs which contains hundreds and thousands of features can be used to get more better results.

\end{document}